\documentclass{article}

\PassOptionsToPackage{numbers, compress}{natbib}
\usepackage[preprint]{neurips_2026}
\usepackage[utf8]{inputenc} % allow utf-8 input
\usepackage[T1]{fontenc}    % use 8-bit T1 fonts
\usepackage{hyperref}       % hyperlinks
\usepackage{url}            % simple URL typesetting
\usepackage{booktabs}       % professional-quality tables
\usepackage{amsfonts}       % blackboard math symbols
\usepackage{nicefrac}       % compact symbols for 1/2, etc.
\usepackage{microtype}      % microtypography
\usepackage{xcolor}         % colors
\usepackage{cleveref}

\usepackage{tikz}
\usetikzlibrary{
  arrows.meta,
  positioning,
  shapes.geometric,
  shapes.symbols,
  fit,
  backgrounds,
  calc,
  decorations.pathreplacing
}

\definecolor{synthetic}{RGB}{74,111,165}
\definecolor{human}{RGB}{165,89,74}
\definecolor{capital}{RGB}{93,130,82}
\definecolor{middle}{RGB}{130,130,130}
\definecolor{lightgraybox}{RGB}{245,245,245}

\tikzset{
  title/.style={font=\Large\bfseries},
  subtitle/.style={font=\normalsize\itshape},
  label/.style={font=\small},
  tinylabel/.style={font=\scriptsize},
  nodebox/.style={
    rounded corners,
    draw=black!45,
    fill=white,
    align=center,
    inner sep=5pt,
    font=\small
  },
  syntheticbox/.style={
    nodebox,
    draw=synthetic!70,
    fill=synthetic!8
  },
  humanbox/.style={
    nodebox,
    draw=human!70,
    fill=human!8
  },
  capitalbox/.style={
    nodebox,
    draw=capital!70,
    fill=capital!8
  },
  middlebox/.style={
    nodebox,
    draw=middle!70,
    fill=middle!8
  },
  arrow/.style={-{Latex[length=3mm]}, thick},
  softarrow/.style={-{Latex[length=2.5mm]}, thick, draw=black!55}
}

\title{Human-Provenance Verification should be Treated as Labor Infrastructure in AI-Saturated Markets}

\makeatletter

\newcommand{\equalcontribmark}{%
  \if@preprint
    \textsuperscript{*}%
  \else
    \if@neuripsfinal
      \textsuperscript{*}%
    \fi
  \fi
}

\newcommand{\equalcontribtext}{%
  \if@preprint
    \begingroup
      \renewcommand{\thefootnote}{*}%
      \footnotetext{Equal contribution.}%
    \endgroup
  \else
    \if@neuripsfinal
      \begingroup
        \renewcommand{\thefootnote}{*}%
        \footnotetext{Equal contribution.}%
      \endgroup
    \fi
  \fi
}

\makeatother

\author{%
    $\textbf{Erin McGurk}^{\dagger\equalcontribmark}$ \quad 
    $\textbf{David Khachaturov}^{\ddagger\equalcontribmark}$ \\
    $^\dagger$Department of Land Economy \quad $^\ddagger$Department of Computer Science and Technology \\
    University of Cambridge
}

\begin{document}

\maketitle
\equalcontribtext

\begin{abstract}
This position paper argues that AI-saturated markets are likely to create Veblen-good premiums, which we term \textit{human-provenance premiums}, for verified human presence, and hence AI governance should treat human-provenance verification as labor infrastructure. Generative and agentic AI systems lower the cost of many standardized cognitive, creative, and coordination tasks, weakening the scarcity premiums that have supported much middle-tier knowledge work. We argue that this pressure may produce an asymmetric barbell-shaped structure of value capture in advanced economies: high-volume synthetic production controlled by owners of AI infrastructure at one pole, and scarce, high-status human labor valued for verified human presence at the other.

We advance three claims. First, AI compresses the value of standardized middle-tier labor by making \textit{good-enough} synthetic substitutes scalable at low marginal cost, hollowing out the middle of the skill distribution currently categorized by knowledge work. Second, this compression reallocates demand for human labor toward work valued for its visible human character. We term this \emph{performative humanity} and distinguish three forms: relational presence labor, aesthetic provenance labor, and accountability labor. Third, as these premiums depend on credible verification, AI governance should treat human-provenance systems as labor infrastructure rather than as luxury authenticity labels.

To evaluate hybrid human-AI work, we propose \textit{constitutive human presence} as the relevant standard: human labor retains premium value when human judgment, attention, accountability, authorship, or relational participation is not incidental to the output but constitutive of what is being purchased. We argue that provenance infrastructure should therefore be portable, privacy-preserving, auditable, and designed to protect worker bargaining power rather than merely certify authenticity for consumers.
\end{abstract}

\section{Introduction}

\textbf{This paper advances the position that in AI-saturated markets, verified human presence will become a scarce and status-bearing attribute of some forms of labor; therefore, AI governance should treat human-provenance verification as labor infrastructure rather than as a luxury authenticity label.} Verified human presence will become a Veblen-like attribute in some labor markets: its demand will increase rather than fall as its price rises due to signaling of scarcity, status, or authenticity~\citep{veblen1899,leibenstein1950}. We do not claim that AI creates the positional value of human labor from nothing; human provenance already carries premiums in artisanal, relational, professional, and luxury markets. Our claim is that generative and agentic AI generalizes and intensifies this logic by flooding cognitive and creative markets with cheap synthetic substitutes, thereby making verified human presence newly salient as a scarce and status-bearing attribute.

The mechanism runs through what we call the compression of the scarcity premium attached to \emph{middle-tier knowledge work}, which is the broad layer of cognitive, creative, and coordination occupations that historically derived bargaining power from scarce human expertise and costly human execution. These occupations are now exposed to competition from systems that generate \emph{good-enough substitutes}: outputs that are not identical to expert human work, and therefore are not \textit{perfect} substitutes, but are adequate for a wide range of purposes, produced at low and falling marginal cost relative to human labor. As good-enough substitutes scale, the scarcity that anchored middle-tier labor value diminishes. If these mechanisms persist, we predict a barbell pattern of value capture: rents accrue at one pole to owners of synthetic-scale infrastructure, while a narrower premium emerges at the other pole for labor whose value depends on verified human provenance.

We use the following concepts to organize the argument. \textbf{AI-driven optimization} refers to the cheapening of standardized cognitive work; 
%empirical studies already show substantial productivity gains in writing-intensive and service settings \citep{noy2023,brynjolfsson2025}.
We introduce the term \textbf{performative humanity} to refer to labor valued because it is visibly and verifiably human rather than inheriting its value from the value of its output. The premium generated by performative humanity is described as the \textbf{human-provenance premium}. We disaggregate performative humanity into three distinct types: relational presence labor, aesthetic provenance labor, and accountability labor; this is explored in~\Cref{sec:performativity} and summarized in~\Cref{tab:human-presence-types}. We use the term \textbf{constitutive human presence} to describe cases in which the buyer’s reason for valuing a good or service depends on a specific human’s judgment, attention, accountability, or relational participation. This standard allows us to analyze human-provenance premiums derived from hybrid human-AI work without a brittle human-versus-machine binary.

\begin{table}[t]
\centering
\caption{Types of human-presence value and their associated verification challenges.}
\label{tab:human-presence-types}
\small
\begin{tabular}{p{0.20\linewidth} p{0.20\linewidth} p{0.27\linewidth} p{0.20\linewidth}}
\toprule
\textbf{Type of labor} & \textbf{Origin of value} & \textbf{Verification problem} & \textbf{Main risk} \\
\midrule
Relational presence &
Chosen attention by a human moral agent &
Was there meaningful human participation? &
Emotional extraction and surveillance \\
\addlinespace
Aesthetic provenance &
Human origin and trace of effort &
Was the origin claim credible? &
Synthetic mimicry and fraud \\
\addlinespace
Accountability &
Human responsibility and liability &
Who bears legal/professional risk? &
Liability laundering \\
\bottomrule
\end{tabular}
\end{table}

The economic implication of our position is that \textit{labor demand need not collapse for labor value to be reorganized}. A growing literature on occupational exposure shows that large language models can affect a broad range of cognitive tasks \citep{eloundou2024,felten2021,ilo2023,imf2024}. The social implication is that \textit{labor solidifies itself as a source of status, identity, and legitimacy beyond its definition as a factor of production}. If the argument developed here plays out, the most consequential effect of AI will not be unemployment but the \textit{emergence of a new hierarchy in which human labor commands a premium because it cannot be optimized away}. Some forms of verified human labor may acquire Veblen-like properties, but only where human-provenance is scarce, visible, and institutionally verifiable. Our claims, as well as the corresponding evidence currently available, are summarized in~\Cref{tab:claim-evidence-status}.

\textbf{Contributions.} This paper makes four contributions. First, it identifies marginal-cost compression of standardized cognitive and creative work as a mechanism by which generative AI weakens the scarcity premium attached to middle-tier knowledge labor. Second, it introduces a taxonomy of human-presence value: relational presence labor, aesthetic provenance labor, and accountability labor. Third, it proposes constitutive human presence as a standard for evaluating hybrid human-AI work, asking whether human judgment, attention, accountability, authorship, or relational participation is incidental to the output or constitutive of what is being purchased. Fourth, it derives implications for AI governance: provenance infrastructure should be treated as labor infrastructure because verification mediates access to human-provenance premiums, and AI systems should be evaluated according to whether they create complementary human tasks or merely substitute for existing judgment at scale.

\section{Theoretical Background}

The theoretical foundation of our argument begins in labor economics. The task-based polarization literature established that automation tends to hollow out routine middle-skill occupations while expanding employment at both extremes of the wage distribution~\citep{autor2013,goos2014}. The skill-biased technological change framework offered the earlier prediction that computerization would raise demand for high-skill workers across sectors by complementing their analytical capabilities~\citep{autor2013}. Generative AI sits within both frameworks. Its distinctive feature is that it competes within the high-skill tier rather than beneath it, targeting the creative, coordinative, and communicative outputs that prior automation left untouched. The result is a pressure that extends the polarization dynamic further up the skill distribution than either framework anticipated.

The information economics literature supplies the mechanism by which this pressure operates. \citet{shapirovarian1999} established that information goods carry near-zero marginal reproduction costs once produced. Generative AI extends low marginal-cost reproduction into low marginal-cost generation of novel-seeming text, images, plans, and other symbolic outputs: the model is a fixed cost, and each additional output generated from it costs almost nothing at the margin. 

The scarcity that sustained the postindustrial middle class lay in the human time required to produce original cognitive work, and the compression of that cost is what drives the restructuring of value we describe. This economic shift does not occur in a vacuum. The political economy literature on platform capitalism documents how ownership of computational infrastructure and distribution interfaces concentrates gains through rent extraction to create durable asymmetries in which those who own the systems capture increasing returns while those who use them receive diminishing shares of the value they generate \citep{srnicek2017,zuboff2019,varoufakis2024,gilbert2024}. 

\newcommand\posscite[1]{\citeauthor{#1}'s \cite{#1}}

What the economics alone cannot explain is why we do not see a uniform collapse of human labor value but a reorganization of its basis (those qualities and signals from which human labor derives a human-provenance premium) accompanied by a redistribution of where in the economy that premium accrues. \posscite{veblen1899} account of conspicuous consumption, formalized by Leibenstein \citep{leibenstein1950} and Bagwell and Bernheim \citep{bagwellbernheim1996}, established that scarcity and visible costliness can themselves become independent sources of demand. Empirical work on provenance confirms that perceived human effort commands measurable premiums even when the output is perceptually identical to a machine-made alternative \citep{fuchs2015,bellaiche2023}. \citet{darby1973} supply a further structural prediction: when (1) the gap between claimed and verifiable quality is large, (2) markets are large enough to sustain fixed certification costs, and (3) buyers can coordinate, formal verification institutions emerge. Under AI saturation all three conditions converge on human-provenance markets simultaneously.

\section{Conditional Barbell Hypothesis}

The postwar middle class in advanced economies rested on an expanding layer of cognitive occupations whose stability depended on human time and judgment remaining non-substitutable factor inputs. That condition gave the median knowledge worker both bargaining power and social position.

\begin{figure}[t]
\centering
\begin{tikzpicture}[x=1cm,y=1cm, yscale=0.5, transform shape=false]

% ---------- Colors ----------
\definecolor{capitalgreen}{RGB}{91,132,82}
\definecolor{humanred}{RGB}{168,86,72}
\definecolor{middlegrey}{RGB}{125,125,125}

% ---------- Styles ----------
\tikzset{
  axislabel/.style={font=\small, align=center},
  regionlabel/.style={font=\small, align=center},
  smallnote/.style={font=\scriptsize, align=center},
  callout/.style={
    rounded corners=2pt,
    draw=black!45,
    fill=white,
    align=center,
    inner sep=5pt,
    font=\small
  },
  arrow/.style={-{Latex[length=2.2mm]}, thin}
}

% ---------- Plot bounds ----------
\coordinate (origin) at (-5.9,-2.8);
\coordinate (xend)   at (6.05,-2.8);
\coordinate (yend)   at (-5.9,4.7);

% ---------- Axes ----------
\draw[->, thick] (origin) -- (yend);
\draw[->, thick] (origin) -- (xend);

% Y-axis label moved closer to axis
\node[axislabel, anchor=south, rotate=90] at (-6,0.95)
{Relative value captured};

% X-axis label moved above category labels
\node[axislabel, anchor=north] at (0,-3.08)
{Position relative to AI commoditization};

% ---------- Background regions ----------
\fill[capitalgreen!7] (-5.55,-2.65) rectangle (-2.55,4.35);
\fill[middlegrey!7]  (-2.55,-2.65) rectangle (2.05,4.35);
\fill[humanred!7]    (2.05,-2.65) rectangle (5.65,4.35);

\draw[dashed, black!25] (-2.55,-2.65) -- (-2.55,4.35);
\draw[dashed, black!25] (2.05,-2.65) -- (2.05,4.35);

% ---------- Mathematical curve ----------
% y = baseline + large infrastructure-rent peak + smaller human-premium peak
\draw[
  black,
  line width=2.2pt,
  line cap=round,
  domain=-5.35:5.45,
  samples=240,
  smooth
]
plot
(
  \x,
  {
    -2.10
    + 5.35*exp(-0.5*((\x+4.25)/0.82)^2)
    + 1.55*exp(-0.5*((\x-3.35)/0.92)^2)
    + 0.20*exp(-0.5*((\x+0.10)/2.90)^2)
  }
);

% ---------- Within-plot labels ----------
\node[smallnote, text width=2.4cm] at (-4.05,-2)
{defended by\\\textbf{scale ownership}};

\node[smallnote, text width=2.5cm] at (-0.25,0)
{least defended:\\\textbf{AI substitutability}};

\node[smallnote, text width=2.5cm] at (3.45,-2)
{defended by\\\textbf{human scarcity}};

\node[smallnote, text width=2.6cm] (tailnote) at (5.35,0.5)
{unverified /\\non-premium\\human labor};

\draw[-{Latex[length=1.8mm]}, thin, black!55]
(tailnote.south) -- (5.35,-1.70);

% ---------- Region labels below x-axis label ----------
\node[regionlabel, text width=3.0cm, anchor=north] at (-4.05,-3.95)
{\textbf{Synthetic-scale control}\\compute, cloud, data, platforms, models};

\node[regionlabel, text width=3.4cm, anchor=north] at (-0.25,-3.95)
{\textbf{Commoditized middle}\\AI-substitutable cognitive and creative output};

\node[regionlabel, text width=3.1cm, anchor=north] at (3.85,-3.95)
{\textbf{Human-\\presence scarcity}\\relationship, provenance, accountability};

\end{tikzpicture}

\caption{The barbell order as a stylized value-capture curve. The horizontal axis is a latent ordering rather than a directly measured variable: it arranges actors by the source of their value defensibility under AI saturation. The left peak represents rents from control over AI infrastructure/synthetic scale; the trough represents commoditized middle-tier cognitive work; the smaller right peak represents premium labor whose value depends on the scarcity of verified human presence. The declining left tail indicates that not all actors in the synthetic-production ecosystem capture infrastructure rents: dependent app-layer firms, integrators, and users of AI systems may benefit from scale without owning the bottleneck. The declining right tail indicates that not all human labor commands a premium: human presence becomes valuable only when it is scarce, verified, and constitutive of the good or service, while unverified or non-premium human labor remains weakly defended.}
\label{fig:barbell-value-curve}
\end{figure}
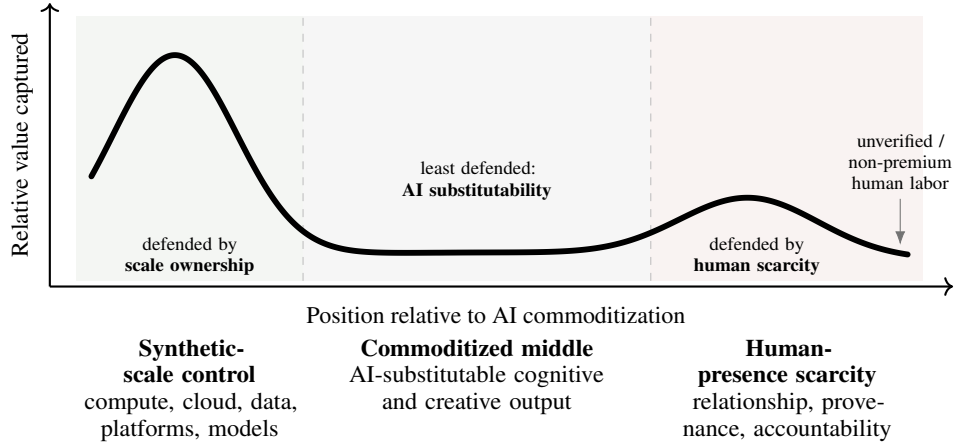

What we see now is less an abrupt breakdown of the postwar settlement than an intensification of a polarization pattern already visible in occupational employment and wage data~\citep{autor2013,goos2014}. Its distinctively AI-era feature is that this polarization pressure now reaches into domains once considered protected by tacit knowledge and symbolic production. \Cref{fig:barbell-value-curve} schematizes the resulting value structure into a \textit{barbell} shape\footnotemark. The left peak represents rents accruing to those who control AI infrastructure and synthetic scale, growing in absolute magnitude as AI deployment expands; the trough represents not a gradual thinning but the structural elimination of the scarcity premium that sustained middle-tier cognitive and creative labor; and the right peak represents a narrow tier of labor whose value depends on verified human presence, high in unit premium but limited in the number of workers it can absorb. The two peaks are asymmetric in both height and growth trajectory: the left captures an expanding share of total value, the right a high but capacity-constrained share. The barbell describes a distribution of economic value, not of employment headcount.

\footnotetext{\textbf{Note:} One can appreciate our policy recommendation without accepting the barbell hypothesis. Even without economy-wide bifurcation, provenance systems still affect worker bargaining power wherever human presence is a priced attribute.}

The most relevant historical precedent runs against the barbell hypothesis -- pre-industrial skilled artisans were not generally premiumized by industrialization; rather, in many trades, factory production competed directly with handicraft production by making the same or similar goods at lower cost, driving down the earnings of hand workers and eventually displacing much of the handicraft sector~\cite{10.1093/ooec/odad033}. The AI case differs structurally as generative AI competes both in the functional economy and in the symbolic economy, where goods are valued not only for what they do but for what they mean and who made them.

In symbolic markets, authorship is a property of the object rather than an incidental fact about its production~\cite{lastowka2005trademark}; AI saturation creates the scarcity conditions for human-provenance value rather than destroying them. That structural asymmetry may not, however, be sufficient to generate a premium tier large enough to absorb the workers displaced from the middle: the market for human provenance may be deep in unit premium while remaining shallow in the number of livelihoods it can sustain; this boundary condition is examined further in Section~\ref{sec:counterarguments}.

\subsection{Why generative and agentic AI change the scarcity structure}

Foundation models convert many forms of cognitive and creative production from labor-time-constrained production into infrastructure-mediated production. Once model access, orchestration, and review workflows are in place, the marginal cost of generating additional deliverables falls substantially relative to human-only production. Agentic systems extend this pressure from isolated outputs to sequences of coordination: monitoring, routing, drafting, scheduling, coding, and customer interaction. The result is a new benchmark against which human labor must justify its price (see~\Cref{fig:labor-regime-comparison} in the Appendix for a historical comparison of labor regimes).

\section{Marginal-Cost Compression of Middle-Tier Work}

Generative models have high fixed development and training costs but low marginal production costs: once a model is trained, additional machine-generated outputs can often be produced at very low, sometimes near-zero, incremental cost~\cite{martens2024economic}. In markets where buyers weigh speed, adequacy, and price over provenance this drives down the baseline against which human labor is priced~\citep{shapirovarian1999,noy2023,brynjolfsson2025}. \citet{noy2023} find in a randomized experiment that access to ChatGPT reduced task completion time and raised assessed output quality in professional writing contexts. \citet{brynjolfsson2025} show that AI assistance raised customer-support productivity substantially, with the largest gains concentrated among less experienced workers, compressing within-occupation skill premiums and thereby collapsing the experience-based wage gradient that sustained middle-tier labor value. Exposure studies confirm that task-level impact extends across clerical, administrative, and professional domains \citep{eloundou2024,felten2021,ilo2023,ilo2025refined}.

As standardized cognitive outputs become cheaper to produce, firms can reduce the headcount required for routine work. The workers who remain need not be fewer in aggregate terms, but they are less scarce, and it is scarcity, not employment per se, that sustains wage premiums and occupational status~\cite{weeden2002occupations,lippmann2008displaced}. The relevant unit of analysis is therefore not the job but the scarcity rent attached to it, and it is the rent, not the role, that AI saturation systematically eliminates.

This mechanism is strongest where outputs are standardized, quality can be cheaply reviewed, buyer tolerance for “good enough” output is high, and provenance is not itself part of the good. It is weaker where errors are costly, tacit domain knowledge is difficult to evaluate, regulation requires human responsibility, or the buyer’s willingness to pay depends on human-provenance~\cite{brynjolfsson2017machine,autor2003skill,bellaiche2023humans}. 

\section{Performative Humanity as a New Source of Value}\label{sec:performativity}

We define \emph{performative\footnote{We use ``performative'' in the sociological sense of being socially enacted and recognized, not to imply superficiality.} humanity} as labor whose economic value depends on its visible human inputs rather than on its output characteristics. A shared insight from the existing literature on emotional labor \citep{hochschild1983}, the sociology of valuation \citep{zelizer1994}, and the economics of credence goods \citep{darby1973} is that ``human labor'' is not a single commodity form. We break it down into three analytically distinct types, each with its own mechanism of value creation and vulnerability to AI substitution.

\textbf{Type I: Relational presence labor.} This is labor whose value is constituted by intersubjectivity. Therapy, care work, tutoring, and other such affective occupations fall into this category. What is being purchased is not primarily an outcome but a relationship: the experience of being known and responded to by another person with their own interiority. The value of a therapeutic relationship, for example, depends on the therapist's status as a moral agent who chooses to attend~\citep{zelizer1994}.

A significant challenge to Type I labor irreplaceability comes from documented formation of emotional bonds between users and AI systems such as Replika, Character.ai, and Woebot \citep{laestadius2024, fitzpatrick2017}. The relevant distinction is between \emph{companionship AI}, which fills a relational vacuum where human alternatives were absent, and \emph{professional relational presence}, in which the practitioner's moral agency, professional accountability, and chosen attendance are constitutive of the value exchanged. The claim is strongest in professional care, therapy, education, and high-trust advisory settings, and weaker in companionship markets where users may accept synthetic substitutes.

Who captures a Type I premium remains a central distributive question. Care work is disproportionately performed by women, migrants, and lower-income workers~\citep{folbre2001,fraser2016}, so AI-driven scarcity may generate a premium without delivering it to those currently performing that labor at scale.

\textbf{Type II: Aesthetic provenance labor.} This is labor valued primarily for its origin narrative. Goods like handmade ceramics and bespoke tailoring carry a premium because the object is understood to encode the trace of a particular person's attention and physical effort. \citet{fuchs2015} document how the attachment of a ``handmade'' label elevates perceived value by invoking embedded human care. \citet{bellaiche2023} similarly show that human authorship labels shift valuations upwards even when the perceived object is identical. \citet{mandel2026} provide incentive-compatible auction evidence that people systematically devalue artwork believed to involve AI, with devaluation following a nonlinear pattern consistent with psychological contamination: even minimal AI involvement produces outsized devaluation, suggesting the human-provenance property is experienced as all-or-nothing rather than as a continuous gradient. This type is the most legible as a Veblen-adjacent phenomenon and the most vulnerable to synthetic mimicry, which is why verification infrastructure becomes load-bearing not just as a consumer authenticity guarantee but also as the mechanism by which aesthetic provenance workers can access and retain the premium their human origin commands; the limits of this guarantee are examined in~\Cref{sec:counterarguments}.

\textbf{Type III: Accountability labor.} This is labor valued due to the worker bearing the risk for the quality of their judgment. Lawyers, physicians, financial advisors, and licensed engineers command premiums for their accountability~\cite{kleiner2000occupational,shapiro1986investment}: they are answerable in ways AI systems currently are not. AI disruption affects this type of labor differently from the others, as the question of replaceability is not contingent on AI matching outputs, but on whether it can assume both legal and moral liability. Given regulatory frameworks requiring a licensed human to bear professional responsibility even when AI is used, accountability labor retains a structural floor that is legally rather than economically maintained~\citep{eu2024aiact}.

These three types of labor often intersect within a single occupation -- a therapist embodies relational presence while bearing legal accountability -- yet they respond to AI pressure through distinct channels~\citep{autor2013task,brynjolfsson2017machine}. Type II is threatened by synthetic mimicry~\citep{bellaiche2023humans,newman2012art}, Type III by regulatory shifts~\citep{freidson2001professionalism,shapiro1986investment}, and Type I by the sheer emotional exhaustion of commodified presence~\citep{hochschild1983,brotheridge2002emotional,hulsheger2011costs}. Analytical conflation of these types risks over-predicting broad premiumization while under-predicting specific vulnerabilities. While Type II and III labor generate new rents for a specialized elite, the premiumization of Type I introduces a new idiom of subordination in which the defensible core of labor increasingly requires the personhood of the worker to be sold as the commodity itself, extending the account of emotional labor developed by~\citet{hochschild1983} into a more stratified AI era~\citep{zelizer2005purchase,boris2010intimate}.

\section{Human Provenance as a Positional Signal}
\label{sec:veblen}

We distinguish a human-provenance premium from a Veblen effect. A human-provenance premium exists when buyers pay more for verified human labor of the kind we term performative humanity than for otherwise comparable synthetic output. A Veblen effect is narrower: it exists when higher price itself increases demand because price signals scarcity, status, or authenticity \citep{veblen1899,leibenstein1950,bagwellbernheim1996}.

Direct evidence of upward-sloping demand for human-made goods in AI-saturated markets has not yet been systematically documented. We therefore frame the micro-level studies cited throughout as establishing the \emph{preconditions} for the Veblen mechanism -- a level-shift in willingness to pay for human-labeled outputs \citep{fuchs2015,bellaiche2023,mandel2026} -- while treating the macro Veblen claim as a forward-looking prediction whose empirical confirmation will depend on observing price-sensitivity patterns as AI saturation deepens. Nascent market evidence is consistent with this mechanism. Cara, a creative portfolio platform prohibiting AI-generated submissions, offers preliminary evidence of market segmentation around human-artist provenance, but not yet evidence of a durable price premium~\cite{cara_about}. \citet{wgaw_ai_2025} similarly made the human-synthetic distinction contractually salient through disclosure requirements and limits on compelled AI use. These developments are consistent with Veblen-type differentiation in which the price premium attaches not to functional superiority but to the conspicuous costliness of human production in the symbolic economy, though their persistence and generalizability remain empirically open questions.

In some markets, human provenance may also acquire Veblen-like properties, but the policy argument does not depend on demonstrating upward-sloping demand: it is sufficient that verification mediates access to a human-provenance premium. The mechanism has a formal analogue in the Spencian signaling tradition; see~\Cref{app:signaling}.

The Veblen mechanism is further stabilized by two compounding structural properties. The first is the \emph{positional} property: \citet{hirsch1976} argued that as material abundance increases, positional competition intensifies, by which it is meant that the social limits on what can be mass-produced are precisely what sustain the premium on what cannot. As synthetic substitutes become abundant, any output that credibly signals the absence of synthesis becomes positional by construction. The second is the \emph{credence} property: human authorship and relational attention are increasingly credence attributes in AI-saturated markets, because buyers cannot reliably determine from the output alone whether a person or a system produced it~\citep{darby1973}. Credence makes verification costly, which raises the price of credibly human labor; the rising price amplifies positional value; this positional intensity feeds back into Veblen demand. All three properties are produced by the same underlying condition (namely, AI saturation) and amplify one another in the same direction.

\section{Economic Implications}
\label{sec:econ_implications}

Control over AI production and distribution infrastructure further concentrates bargaining power at the capital-owning apex of the barbell. Large AI systems depend on a vertically integrated stack of computational and distribution infrastructure; ownership over these layers confers a form of command that exceeds ordinary firm competition because ordinary competition assumes that rivals can contest the same market on comparable terms~\citep{srnicek2017,zuboff2019,varoufakis2024,gilbert2024}. Human-presence workers whose market is structurally limited to those who captured infrastructure rents face a positional-good dynamic: the human-provenance premium is affordable only to those whose purchasing power derives from the same concentration it rewards. 

The immediate economic implication is stronger bifurcation in wages and bargaining power. Workers whose tasks are easily standardized or reviewed by AI will face intensified competition and weaker claims to premium compensation. Workers close to infrastructure ownership or system integration may capture outsized gains. In between lies an unstable zone of downgraded professional labor which remains employed, but is more weakly defended as a source of middle-class security \citep{autor2013,goos2014}.

The Acemoglu-Restrepo ``so-so automation'' thesis \citep{acemoglu2020} argues that automation which substitutes for labor without creating new complementary tasks reduces labor's aggregate income share. If this applies to generative AI, it raises a structural challenge for the barbell hypothesis: a declining aggregate labor income share shrinks the consumer base that could sustain a premium human labor market. The strongest demand for premium human labor is therefore likely to come from two sources: high-income buyers who use human provenance positionally~\citep{veblen1899,bellaiche2023humans,newman2012art}, and institutions that require human accountability for legal, ethical, or legitimacy reasons~\citep{shapiro1986investment,eu2024aiact,freidson2001professionalism,unesco2021ai}. The Veblen market for human authenticity is therefore funded not by a rising labor tier but by the beneficiaries of the capital share itself. For the compressed middle, this tier is structurally inaccessible, priced out by the same redistribution of aggregate output that finances demand at the top.

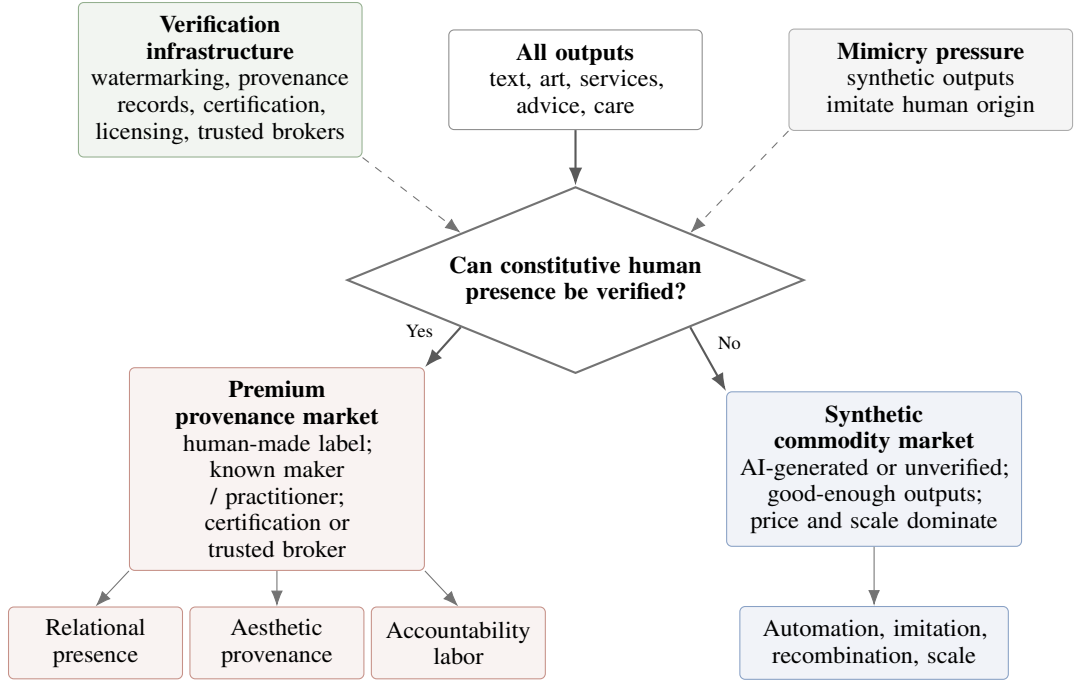
\begin{figure}[t]
\centering
\begin{tikzpicture}[x=1cm,y=1cm]

% ---------- Colors ----------
\definecolor{humanred}{RGB}{168,86,72}
\definecolor{syntheticblue}{RGB}{74,111,165}
\definecolor{capitalgreen}{RGB}{91,132,82}
\definecolor{middlegrey}{RGB}{125,125,125}

% ---------- Styles ----------
\tikzset{
  nodebox/.style={
    rounded corners=2pt,
    draw=black!45,
    fill=white,
    align=center,
    inner sep=5pt,
    font=\small
  },
  humanbox/.style={
    nodebox,
    draw=humanred!65,
    fill=humanred!7
  },
  syntheticbox/.style={
    nodebox,
    draw=syntheticblue!65,
    fill=syntheticblue!8
  },
  greenbox/.style={
    nodebox,
    draw=capitalgreen!65,
    fill=capitalgreen!8
  },
  greybox/.style={
    nodebox,
    draw=middlegrey!65,
    fill=middlegrey!8
  },
  arrow/.style={-{Latex[length=2.4mm]}, thick, draw=black!65},
  softarrow/.style={-{Latex[length=2.1mm]}, thin, draw=black!55},
  smalllabel/.style={font=\scriptsize, align=center},
  influence/.style={-{Latex[length=2.1mm]}, dashed, thin, draw=black!55},
  smallhumanbox/.style={
    humanbox,
    text width=1.95cm,
    minimum width=2.25cm,
    minimum height=0.95cm,
    inner ysep=4pt,
    align=center
  }
}

% ---------- Upstream inputs ----------
\node[greenbox, text width=3.4cm] (infra) at (-4.7,4.8)
{\textbf{Verification\\infrastructure}\\watermarking, provenance records, certification, licensing, trusted brokers};

\node[nodebox, text width=3.0cm] (all) at (0,4.8)
{\textbf{All outputs}\\text, art, services, advice, care};

\node[greybox, text width=3.4cm] (mimicry) at (4.7,4.8)
{\textbf{Mimicry pressure}\\synthetic outputs imitate human origin};

% ---------- Gate ----------
\node[
  diamond,
  aspect=2.45,
  draw=black!55,
  fill=white,
  thick,
  align=center,
  inner sep=1pt,
  text width=4.25cm,
  font=\small
] (gate) at (0,2.15)
{\textbf{Can constitutive human\\presence be verified?}};

\draw[influence] (infra.south east) -- (gate.north west);
\draw[influence] (mimicry.south west) -- (gate.north east);
\draw[arrow] (all.south) -- (gate.north);

% ---------- Outcomes ----------
\node[humanbox, text width=3.55cm] (premium) at (-3.95,-0.35)
{\textbf{Premium\\provenance market}\\human-made label;\\known maker / practitioner;\\certification or trusted broker};

\node[syntheticbox, text width=3.55cm] (commodity) at (3.95,-0.35)
{\textbf{Synthetic\\commodity market}\\AI-generated or unverified;\\good-enough outputs;\\price and scale dominate};

\draw[arrow] (gate.south west) -- node[above left, smalllabel] {Yes} (premium.north east);
\draw[arrow] (gate.south east) -- node[above right, smalllabel] {No} (commodity.north west);

% ---------- Premium mechanisms, spaced row ----------
\node[smallhumanbox] (rel) at (-6.35,-2.65)
{Relational presence};

\node[smallhumanbox] (prov) at (-3.95,-2.65)
{Aesthetic provenance};

\node[smallhumanbox] (acct) at (-1.55,-2.65)
{Accountability labor};

\draw[softarrow] (premium.south west) -- (rel.north);
\draw[softarrow] (premium.south) -- (prov.north);
\draw[softarrow] (premium.south east) -- (acct.north);

% ---------- Synthetic pathway ----------
\node[syntheticbox, text width=3.2cm] (auto) at (3.95,-2.65)
{Automation, imitation, recombination, scale};

\draw[softarrow] (commodity.south) -- (auto.north);

\end{tikzpicture}

\caption{The verification gate separating premium human-provenance markets from synthetic commodity markets. Verification infrastructure and mimicry pressure exert opposing pressures on the gate: the former makes human presence easier to certify, while the latter makes authentication more difficult. The resulting market split depends on whether human judgment, attention, and accountability can be verified as constitutive of the good or service.}
\label{fig:verification_gate}
\end{figure}

Aggregate productivity may increase even as workers experience declining status and bargaining power~\citep{lippmann2008displaced,acemoglu2019automation,leduc2024automation}. At the consumer level, one should expect deeper segmentation where mass markets are dominated by synthetic low-cost goods and services~\citep{martens2024economic,kirk2025ai}, and premium markets increasingly bundle human provenance with exclusivity~\citep{veblen1899,bellaiche2023humans,newman2012art}. \Cref{fig:verification_gate} schematizes this segmentation as a verification gate: outputs whose constitutive human presence can be verified enter premium human-provenance markets, while synthetic or unverified outputs are routed into commodity markets. Institutional differences will shape the pace: the EU's AI Act may moderate speed and some harms \citep{eu2024aiact}, while more market-led frameworks may produce faster and sharper bifurcation \citep{nist2023}; the underlying pressure toward concentration appears common across settings~\citep{ilo2023,imf2024}.

\section{Provenance Design Requirements for Labor Infrastructure}

If human provenance is to function as labor infrastructure, the relevant technical question extends beyond detection of AI-generated content to whether provenance systems preserve worker bargaining power, accountability, privacy, and market access under conditions of hybrid production. C2PA-style Content Credentials, for example, can record asset history, modifications, and AI use in cryptographically bound provenance records~\citep{c2pa2024}. Yet such records do not by themselves establish whether a worker had \emph{constitutive human presence}. Hence, labor-facing provenance therefore requires five additional design principles.

\textbf{1. Process provenance, not only content provenance.} Labor provenance must represent what happened in the work process. The relevant attribute is \emph{constitutive human presence}, and provenance systems should therefore support attestations about the human role in production rather than reducing verification to output classification.

\textbf{2. Human-role attestations for hybrid work.} Hybrid human-AI production will likely be the ordinary case, hence provenance systems should specify who exercised final judgment, who bore liability, who supplied relational labor, and which stages of production were synthetic. A labor-infrastructural system should at minimum distinguish human authors, human reviewers, accountable parties, and relational participants. These categories map onto the three forms of performative humanity; see~\Cref{tab:human-presence-types}.

\textbf{3. Privacy-preserving worker verification.} Verification can easily become surveillance if it relies on continuous process monitoring rather than targeted attestation. Worker-facing provenance should favor privacy-preserving attestations that confirm constitutive human presence without requiring continuous monitoring of how that presence was exercised.

\textbf{4. Portability across platforms.} If platforms control the infrastructure through which human provenance is established and made legible to buyers, verification may deepen platform dependency rather than reduce it. Workers could lose accumulated provenance when leaving a platform, while intermediaries capture the human-provenance premium through certification and discovery. We hence advocate for labor-facing provenance to be portable across platforms and professional institutions. Portability is a labor-right requirement as it determines whether provenance increases worker bargaining power or creates a new bottleneck.

\textbf{5. Auditability of authenticity scoring.} Where human provenance becomes economically valuable, authenticity scoring systems will emerge to rank and price workers by perceived humanness. These systems inherit the well-documented failure modes of algorithmic management: discrimination, gaming, false positives, and the enforcement of platform-legible behavioral norms under the guise of quality measurement. Treating them as high-stakes labor-market infrastructure requires mandatory audits that ask not only whether scores are accurate but whose interests the scoring criteria serve.

These principles also imply a negative requirement: provenance \textit{should not} be collapsed into AI-output detection. Reliable AI-text detection~\citep{sadasivan2024detection} and watermarking~\citep{khachaturov2025watermarking} remain technically contested. NIST accordingly frames the above -- provenance, detection, auditing, etc. -- as a broader technical ecosystem rather than a solved problem~\citep{nist2024synthetic}.

\section{Counterarguments and Objections}
\label{sec:counterarguments}

The barbell hypothesis is posed as a conditional prediction rather than a claim of inevitability. Institutional ruptures -- collective labor action, antitrust intervention, publicly-owned AI infrastructure, or stronger labor regulation such as the recent Chinese court ruling restricting firms from laying workers off on AI ground~\cite{xu2026aihangzhoulabor} -- could redirect the trajectory described here. The objections below clarify the conditions under which the argument holds.

\textbf{Objection 1: AI will complement rather than hollow out labor.} Studies find productivity gains and especially strong effects for less experienced workers, which could reduce within-occupation inequality~\citep{noy2023,brynjolfsson2025,ilo2023,oecd2025sme}. We agree that complementarity is possible. Our narrower claim is that task-level complementarity does not preclude occupational devaluation. A worker can become more productive and yet less scarce if many others, aided by AI, can produce similar outputs. Productivity gains may therefore coexist with declining labor bargaining power when the scarcity of specialized expertise is structurally eroded.

\textbf{Objection 2: Authenticity can be simulated.} If synthetic goods can be marketed convincingly as artisanal, the premium on human provenance should collapse. This is a serious challenge: if authenticity is not verifiable, it cannot reliably be purchased. We resolve the tension by distinguishing two stages of authentication. In the first, authentication is social: elites verify human-provenance through proximity, direct commissioning, and known makers. In the second, as synthetic mimicry improves and social verification proves too narrow, institutional verification markets emerge through certification bodies, watermarking regimes, human-made labeling standards, and trusted brokers~\citep{bellaiche2023,eu2024aiact}. The contradiction is therefore temporal rather than fatal: informal verification precedes formal verification, and the latter becomes necessary as mimicry scales.

\textbf{Objection 3: Verification can become surveillance.} Human-provenance systems may require intrusive monitoring of workers' creative process, attention, affect, or tool use. A system that verifies human presence through continuous monitoring would preserve authenticity only by deepening managerial control. A labor-infrastructural approach should therefore distinguish verification from surveillance. Provenance systems should be designed around the minimum process record necessary to establish the relevant labor claim, and should treat affective monitoring as categorically incompatible with certification.

\textbf{Objection 4: The premium may be exclusionary or captured by platforms.} Even if human provenance becomes valuable, platforms may capture the premium by controlling verification, ranking, and access to buyers. The existence of a human-provenance premium does not imply that workers receive it; platform labor markets already show that human labor does not automatically become premium when mediated by digital systems, since platforms often increase substitutability and compress wages in routinizable services~\citep{srnicek2017}. This objection strengthens the paper's central claim that provenance must be treated as labor infrastructure, not as basic consumer-facing authenticity labeling. Worker-protective provenance systems should require portability across platforms, auditability of ranking and authenticity-scoring systems, and collective governance by the workers whose labor is being certified. Furthermore, workers in informal, low-income, Global South, migrant, or care economies may be least able to afford certification, even when their labor is most human-dependent. This connects directly to the distributional concern that relational presence and care labor are often gendered, racialized, migrantized, and historically undervalued~\citep{folbre2001,fraser2016}. Access to verification must accordingly be treated as a distributive question, with standards designed with marginalized labor groups rather than imposed on them after the fact.

\section{Conclusion}

This paper argues that AI saturation reorganizes the social basis on which labor retains value, making human-provenance verification a problem of labor infrastructure rather than as a luxury authenticity label. The central question for AI governance is therefore not whether human labor will survive automation, but which forms of human presence will remain economically, legally, and socially defensible. If provenance infrastructure is built only for premium consumer markets, AI will likely intensify the \textit{barbell} hypothesis described here. If it is built as labor infrastructure, it may instead preserve bargaining power for workers whose judgment, accountability, care, and authorship are constitutive of the value they create.

The barbell hypothesis remains a conditional prediction. Two constraints are especially important. First, the human-premium tier may be deep in value but shallow in employment capacity, redistributing value without absorbing workers displaced from the middle. Second, the workers most likely to supply relational presence and care are often those least positioned to capture the resulting premium, given the gendered and racialized organization of care labor~\citep{folbre2001,fraser2016}. Any policy account that ignores these constraints will overstate the welfare gains from human-provenance markets.

\textbf{Resulting research agenda.} Provenance infrastructure should be designed as labor infrastructure: portable across platforms, auditable, privacy-preserving, and capable of representing human role, consent, accountability, and compensation. Systems in high-exposure occupations should also be evaluated for whether they create complementary human tasks or merely substitute for existing judgment~\citep{acemoglu2020}. Authenticity-scoring systems require audit to ensure that they measure genuine service quality rather than enforcing platform-legible norms of affective performance~\citep{hochschild1983}. Finally, workers in the three human-presence categories should participate directly in the design of systems that verify, rank, or price their labor.

\bibliographystyle{unsrtnat}
\bibliography{references}

%%%%%%%%%%%%%%%%%%%%%%%%%%%%%%%%%%%%%%%%%%%%%%%%%%%%%%%%%%%%

\newpage

\appendix
\section{Summary of paper's core claims}

\begin{table}[h]
\centering
\caption{Evidentiary status of the paper's core claims.}
\label[appendix]{tab:claim-evidence-status}
\small
\begin{tabular}{p{0.30\linewidth} p{0.36\linewidth} p{0.26\linewidth}}
\toprule
\textbf{Claim} & \textbf{Evidence currently available} & \textbf{Status} \\
\midrule
AI reduces the cost of standardized cognitive work &
Task-level productivity studies and information-goods cost theory~\citep{noy2023,brynjolfsson2025,shapirovarian1999} &
Supported as mechanism \\
\addlinespace
Middle-tier labor loses scarcity premium &
Occupational exposure studies, polarization literature, and automation theory~\citep{autor2013,goos2014,eloundou2024,felten2021,ilo2023,ilo2025refined} &
Plausible but not directly demonstrated \\
\addlinespace
Human-origin labels raise willingness to pay &
Provenance, handmade-effect, and AI-art valuation studies~\citep{fuchs2015,bellaiche2023,mandel2026} &
Supported in bounded domains \\
\addlinespace
Human labor becomes Veblen-like &
No direct systematic evidence yet; supported mainly by Veblen, positional-goods, and signaling theory~\citep{veblen1899,leibenstein1950,bagwellbernheim1996,hirsch1976,spence1973} &
Forward-looking prediction \\
\addlinespace
Provenance infrastructure affects worker bargaining power &
Certification economics, credence-goods theory, AI governance rules, and WGA-style labor-contract examples~\citep{darby1973,spence1973,eu2024aiact,wgaw_ai_2025}  &
Normative and under-tested \\
\bottomrule
\end{tabular}
\end{table}

\section{Spencian account of human-provenance as a separating signal}
\label[appendix]{app:signaling}

The two-stage Veblen signaling mechanism described in Section~\ref{sec:veblen} has a formal analogue in the Spencian tradition \citep{spence1973}. This appendix sketches the separating equilibrium under which human-provenance functions as a credible signal.

\paragraph{Setup.} Let there be two types of goods: \emph{human-made} ($H$) and \emph{AI-made} ($A$). Both can carry a human-made label ($s = 1$) or an unlabeled or AI-labeled signal ($s = 0$). Let $\pi > 0$ denote the premium buyers pay for a credibly human-labeled good in an AI-saturated market. Let $c_A > 0$ denote the cost borne by an AI producer of falsely claiming human origin---including reputational exposure and legal liability. Let $c_H \geq 0$ denote the cost borne by a human producer of credibly documenting genuine human origin (through such verifications as provenance records, certification, and watermarking).

\paragraph{Separating equilibrium.} A separating equilibrium exists when:
\begin{equation}
  c_A > \pi > c_H. \label{eq:sep}
\end{equation}
Under condition (\ref{eq:sep}), human producers find it worthwhile to acquire the provenance signal (since $\pi > c_H$), while AI producers find false certification unprofitable (since $c_A > \pi$). The signal is therefore informative, and the premium is sustained in equilibrium.

\paragraph{Dynamics under AI saturation.} Two forces operate as AI saturation deepens. First, $\pi$ increases: as synthetic outputs become more abundant and competent, human provenance becomes more valuable as a status and authenticity signal. Second, $c_A$ is placed under downward pressure as mimicry improves. The separating equilibrium is maintained only if provenance infrastructure keeps $c_A$ above $\pi$---only if verification technology keeps pace with mimicry technology. The paper's recommendation to develop provenance infrastructure is therefore a structural condition for the equilibrium's stability.

\paragraph{Relation to the Veblen mechanism.} The human-provenance signal differs from standard Spencian signals in that its cost structure is asymmetric by construction: AI-produced goods cannot acquire the human-made signal without defeating the claim itself. This asymmetry means the separating condition in (\ref{eq:sep}) is partially guaranteed by the logical structure of the provenance claim. As AI saturation raises $\pi$, the status-signaling value of the human-made label intensifies -- consistent with the Veblen property that rising price increases rather than suppresses demand among status-seeking buyers.

\section{Historical shapes of labor regimes}

\begin{figure}[h]
\centering
\resizebox{\textwidth}{!}{%
\begin{tikzpicture}[
    x=1cm,y=1cm,
    >=Latex,
    title/.style={font=\bfseries\large, align=center},
    small/.style={font=\scriptsize, align=center},
    tiny/.style={font=\tiny, align=center},
    note/.style={font=\scriptsize, align=center, text width=4.05cm},
    panelbox/.style={draw=black!8, rounded corners=3pt, fill=black!1},
    softarrow/.style={-{Latex[length=2.1mm]}, line width=0.6pt, draw=black!55},
    link/.style={line width=0.55pt, draw=black!35},
    capitalbox/.style={
        draw=capital!65, fill=capital!10, rounded corners=5pt,
        minimum height=0.52cm, align=center, font=\scriptsize
    },
    humanbox/.style={
        draw=human!65, fill=human!10, rounded corners=5pt,
        minimum height=0.52cm, align=center, font=\scriptsize
    },
    middlebox/.style={
        draw=black!35, fill=black!4, rounded corners=5pt,
        minimum height=0.52cm, align=center, font=\scriptsize
    },
    landbox/.style={
        draw=brown!55!black, fill=brown!10, rounded corners=4pt,
        minimum height=0.48cm, align=center, font=\scriptsize
    }
]

% ------------------------------------------------------------------
% Uncomment these if the colors are not already defined elsewhere.
% \definecolor{capital}{RGB}{96,140,96}
% \definecolor{human}{RGB}{190,105,92}
% \definecolor{middle}{RGB}{145,145,145}
% \definecolor{synthetic}{RGB}{94,135,190}
% \definecolor{lightgraybox}{RGB}{247,247,247}
% ------------------------------------------------------------------

% Panel centers
\coordinate (HG) at (-5.8,  4.2);
\coordinate (ME) at ( 0.0,  4.2);
\coordinate (RS) at ( 5.8,  4.2);
\coordinate (IP) at (-5.8, -2.0);
\coordinate (PI) at ( 0.0, -2.0);
\coordinate (AI) at ( 5.8, -2.0);

% Soft panel backgrounds
\foreach \C in {HG,ME,RS,IP,PI,AI} {
  \node[panelbox, minimum width=5.0cm, minimum height=5.35cm] at (\C) {};
}

% ================================================================
% Hunter-gatherer: kin-band network
% ================================================================
\node[title] at ($(HG)+(0,2.25)$) {Hunter-gatherer};

\coordinate (hg1) at ($(HG)+(-1.25,0.75)$);
\coordinate (hg2) at ($(HG)+(0.00,1.02)$);
\coordinate (hg3) at ($(HG)+(1.25,0.75)$);
\coordinate (hg4) at ($(HG)+(-0.82,-0.35)$);
\coordinate (hg5) at ($(HG)+(0.82,-0.35)$);

\draw[link] (hg1) -- (hg2);
\draw[link] (hg2) -- (hg3);
\draw[link] (hg1) -- (hg4);
\draw[link] (hg2) -- (hg4);
\draw[link] (hg2) -- (hg5);
\draw[link] (hg3) -- (hg5);
\draw[link] (hg4) -- (hg5);
\draw[link] (hg1) -- (hg5);
\draw[link] (hg3) -- (hg4);

\foreach \p in {hg1,hg2,hg3,hg4,hg5} {
  \draw[thick, fill=capital!7] (\p) circle (0.54);
}

\node[note] at ($(HG)+(0,-1.95)$)
{\textbf{Kin-band network}\\scarcity: ecology, skill, reciprocity};

% ================================================================
% Medieval England: estate hierarchy
% ================================================================
\node[title] at ($(ME)+(0,2.25)$) {Medieval England};

\draw[thick, fill=human!8]
  ($(ME)+(-1.45,-0.75)$) -- ($(ME)+(0,1.55)$) -- ($(ME)+(1.45,-0.75)$) -- cycle;

\draw[black!45] ($(ME)+(-1.02,-0.02)$) -- ($(ME)+(1.02,-0.02)$);
\draw[black!45] ($(ME)+(-0.58,0.70)$) -- ($(ME)+(0.58,0.70)$);

\node[small] at ($(ME)+(0,0.97)$) {King};
\node[small] at ($(ME)+(0,0.24)$) {Lords / clergy};
\node[small] at ($(ME)+(0,-0.40)$) {Peasants};

\node[note] at ($(ME)+(0,-1.95)$)
{\textbf{Estate hierarchy}\\scarcity: land, rank, obligation};

% ================================================================
% Russian serfdom: bound estate order
% ================================================================
\node[title] at ($(RS)+(0,2.25)$) {Russian serfdom};

\node[capitalbox, text width=2.55cm] (tsar) at ($(RS)+(0,1.38)$)
{Tsar / state};

\node[humanbox, text width=2.75cm] (nobility) at ($(RS)+(0,0.48)$)
{nobility / landlords};

\node[landbox, text width=3.05cm] (estate) at ($(RS)+(0,-0.32)$)
{estate / commune};

\node[middlebox, text width=2.75cm] (serfs) at ($(RS)+(0,-1.18)$)
{serf households\\tied to land};

\draw[softarrow] (tsar) -- (nobility);
\draw[softarrow] (nobility) -- (estate);
\draw[softarrow] (estate) -- (serfs);

\node[tiny, anchor=west] at ($(RS)+(0.32,0.9)$) {grant / service};
\node[tiny, anchor=west] at ($(RS)+(0.32,0.00)$) {jurisdiction};
\node[tiny, anchor=west] at ($(RS)+(0.32,-0.7)$) {labor dues};

\node[note] at ($(RS)+(0,-2.08)$)
{\textbf{Bound estate order}\\scarcity: land control, restricted mobility, coerced labor};

% ================================================================
% Industrial / postwar: broad middle order -- more room below
% ================================================================
\node[title] at ($(IP)+(0,2.25)$) {Industrial / postwar};

\draw[thick, fill=lightgraybox]
  ($(IP)+(0,1.18)$) --
  ($(IP)+(-1.28,0.38)$) --
  ($(IP)+(-1.05,-0.92)$) --
  ($(IP)+(0,-1.42)$) --
  ($(IP)+(1.05,-0.92)$) --
  ($(IP)+(1.28,0.38)$) -- cycle;

\draw[black!35] ($(IP)+(-0.58,0.66)$) -- ($(IP)+(0.58,0.66)$);
\draw[black!35] ($(IP)+(-0.98,-0.03)$) -- ($(IP)+(0.98,-0.03)$);
\draw[black!35] ($(IP)+(-0.70,-0.70)$) -- ($(IP)+(0.70,-0.70)$);

\node[small] at ($(IP)+(0,0.80)$) {small elite};
\node[small] at ($(IP)+(0,0.28)$) {\textbf{organized}\\\textbf{middle}};
\node[small] at ($(IP)+(0,-0.47)$) {working class};

\node[note] at ($(IP)+(0,-2.10)$)
{\textbf{Broad middle order}\\scarcity: factories,\\execution, organizations};

% ================================================================
% Post-industrial: hourglass structure
% ================================================================
\node[title] at ($(PI)+(0,2.25)$) {Post-industrial};

% More visibly hourglass-shaped
\draw[thick, fill=lightgraybox]
  ($(PI)+(-1.35,1.10)$) --
  ($(PI)+(1.35,1.10)$) --
  ($(PI)+(0.58,0.10)$) --
  ($(PI)+(1.05,-1.25)$) --
  ($(PI)+(0,-1.75)$) --
  ($(PI)+(-1.05,-1.25)$) --
  ($(PI)+(-0.58,0.10)$) -- cycle;

\draw[black!35] ($(PI)+(-1.00,0.62)$) -- ($(PI)+(1.00,0.62)$);
\draw[black!35] ($(PI)+(-0.50,0.02)$) -- ($(PI)+(0.50,0.02)$);
\draw[black!35] ($(PI)+(-0.80,-0.88)$) -- ($(PI)+(0.80,-0.88)$);

\node[small] at ($(PI)+(0,0.86)$) {finance / tech / owners};
\node[small] at ($(PI)+(0,0.32)$) {credentialed\\work};
\node[small] at ($(PI)+(0,-0.47)$) {services / gig};
\node[small] at ($(PI)+(0,-1.22)$) {low-wage\\service mass};

\node[note] at ($(PI)+(0,-2.10)$)
{\textbf{Hourglass structure}\\scarcity: credentials, networks, capital};

% ================================================================
% AI barbell: horizontal value order
% ================================================================
\node[title] at ($(AI)+(0,2.25)$) {AI barbell};

% Two value poles
\fill[capital!18, draw=capital!70, thick]
  ($(AI)+(-1.18,0.62)$) ellipse (0.88 and 0.46);
\fill[human!16, draw=human!70, thick]
  ($(AI)+(1.18,0.62)$) ellipse (0.88 and 0.46);

% Compressed middle trough
\fill[middle!12, draw=middle!60, thick]
  ($(AI)+(0,0.02)$) ellipse (0.58 and 0.34);

% Synthetic mass market base
\fill[synthetic!18, draw=synthetic!70, thick]
  ($(AI)+(0,-1.05)$) ellipse (1.68 and 0.48);

% Light barbell connector
\draw[black!35, line width=1.2pt]
  ($(AI)+(-0.30,0.62)$) -- ($(AI)+(0.30,0.62)$);
\draw[black!35, dashed]
  ($(AI)+(0,0.60)$) -- ($(AI)+(0,0.22)$);

\node[small] at ($(AI)+(-1.18,0.62)$) {infrastructure\\rents};
\node[small] at ($(AI)+(1.18,0.62)$) {human\\premium};
\node[small] at ($(AI)+(0,0.02)$) {thin\\middle};
\node[small] at ($(AI)+(0,-1.05)$) {synthetic\\mass market};

\node[note] at ($(AI)+(0,-2.10)$)
{\textbf{Barbell value order}\\scarcity: compute\\control and human provenance};

\end{tikzpicture}%
}
\caption{Historical shapes of labor regimes. The panels schematically compare how different economic orders organize scarcity, status, and dependence: hunter-gatherer bands through kin-based reciprocity and ecological skill; medieval hierarchy through land, rank, and obligation; Russian serfdom through bound estate dependency, restricted mobility, and coerced agrarian labor; industrial/postwar capitalism through a broad middle supported by scarce human execution and organizational membership; post-industrial capitalism through an hourglass structure of capital ownership, credentials, and service work; and the hypothesized AI barbell through infrastructure rents, verified human presence, a compressed middle, and synthetic mass consumption. These are conceptual schematics rather than quantitative distributions.}
\label{fig:labor-regime-comparison}
\end{figure}
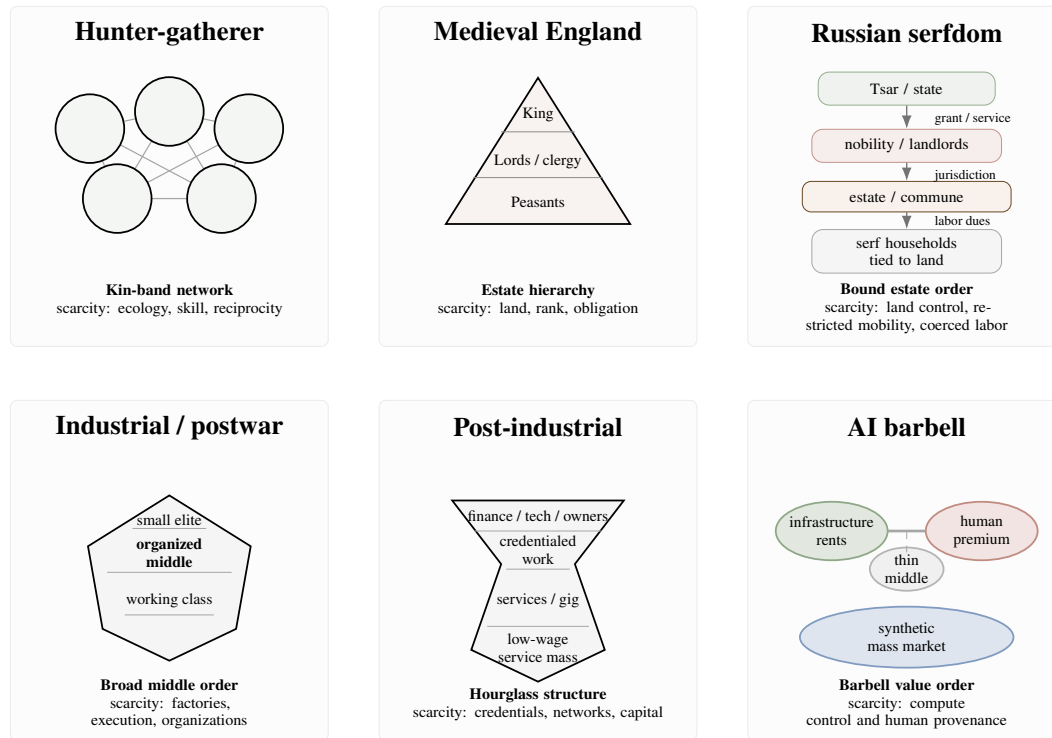

\newpage

\section{Visualizing the transition into a barbell economy}

\begin{figure}[h]
\centering
\begin{tikzpicture}[x=1cm,y=1cm]

\definecolor{capitalgreen}{RGB}{91,132,82}
\definecolor{humanred}{RGB}{168,86,72}
\definecolor{syntheticblue}{RGB}{74,111,165}
\definecolor{middlegrey}{RGB}{125,125,125}

% ---------- Styles ----------
\tikzset{
  titlelabel/.style={font=\bfseries\large, align=center},
  tierlabel/.style={font=\small, align=center},
  smallnote/.style={font=\scriptsize, align=center},
  arrow/.style={-{Latex[length=2.5mm]}, thick},
  softline/.style={draw=black!45, thin},
  tierbox/.style={
    draw=black!60,
    fill=white,
    rounded corners=1.5pt,
    align=center,
    inner sep=4pt,
    font=\small
  }
}

% Left panel title
\node[font=\large\bfseries] at (-4.4,4.75) {Broad middle economy};

% Diamond layers
\coordinate (Ltop) at (-4.4,4.35);
\coordinate (Lupper) at (-5.9,2.65);
\coordinate (Lmidleft) at (-6.75,0.75);
\coordinate (Lbot) at (-4.4,-2.35);
\coordinate (Lmidright) at (-2.05,0.75);
\coordinate (Lupperright) at (-2.9,2.65);

\draw[thick, fill=lightgraybox] 
  (Ltop) -- (Lupperright) -- (Lmidright) -- (Lbot) -- (Lmidleft) -- (Lupper) -- cycle;

% layer separators
\draw[black!35] (-5.25,3.35) -- (-3.55,3.35);
\draw[black!35] (-6.15,1.9) -- (-2.65,1.9);
\draw[black!35] (-6.4,0.35) -- (-2.4,0.35);
\draw[black!35] (-5.3,-1.05) -- (-3.5,-1.05);

\node[label] at (-4.4,3.7) {Elites};
\node[label] at (-4.4,2.55) {Upper-middle};
\node[label, align=center] at (-4.4,1.1) {\textbf{Broad middle class}\\industrial, managerial,\\white-collar labor};
\node[label] at (-4.4,-0.45) {Working class};
\node[label] at (-4.4,-1.65) {Poor};

% Arrow
\draw[arrow] (-1.35,1.0) -- (1.35,1.0)
node[midway, above, label] {AI saturation}
node[midway, below, tinylabel] {marginal-cost collapse};

\node[titlelabel] at (4.7,4.75) {AI barbell economy};

% --- Top tier: infrastructure owners ---
\filldraw[
  draw=capitalgreen!75,
  fill=capitalgreen!12,
  thick
] (4.7,3.35) ellipse (2.25 and 0.65);

\node[tierlabel, text width=3.85cm] at (4.7,3.35)
{\textbf{Infrastructure owners}\\compute, models, platforms};

% --- Premium human tier ---
\filldraw[
  draw=humanred!75,
  fill=humanred!10,
  thick
] (4.7,1.70) ellipse (2.25 and 0.70);

\node[tierlabel, text width=4.05cm] at (4.7,1.65)
{\textbf{Verified human premium}\\presence, provenance, accountability};

% --- Compressed middle ---
\filldraw[
  draw=middlegrey!65,
  fill=white,
  thick
] (4.7,0.10) ellipse (0.92 and 0.42);

\node[smallnote, text width=1.55cm] at (4.7,0.10)
{compressed\\middle};

% --- Synthetic mass market ---
\filldraw[
  draw=syntheticblue!75,
  fill=syntheticblue!12,
  thick
] (4.7,-1.75) ellipse (2.65 and 0.82);

\node[tierlabel, text width=4.65cm] at (4.7,-1.75)
{\textbf{Synthetic mass market}\\cheap goods, media, services};

% --- Simple vertical flow arrows ---
\draw[-{Latex[length=2mm]}, thin, black!45] (4.7,2.68) -- (4.7,2.42);
\draw[-{Latex[length=2mm]}, thin, black!45] (4.7,1.00) -- (4.7,0.56);
\draw[-{Latex[length=2mm]}, thin, black!45] (4.7,-0.34) -- (4.7,-0.90);

% --- Optional light hourglass outline, behind the nodes ---
\begin{scope}[on background layer]
  \draw[black!25, line width=0.9pt]
    (2.95,3.35)
    .. controls (3.45,2.35) and (4.15,1.00) .. (4.45,0.10)
    .. controls (4.05,-0.50) and (3.10,-1.05) .. (2.20,-1.75);

  \draw[black!25, line width=0.9pt]
    (6.45,3.35)
    .. controls (5.95,2.35) and (5.25,1.00) .. (4.95,0.10)
    .. controls (5.35,-0.50) and (6.30,-1.05) .. (7.20,-1.75);
\end{scope}

\end{tikzpicture}
\caption{Schematic transition from a broad middle-class economy to an AI-era barbell structure. The left panel depicts a diamond-shaped distribution in which value and bargaining power are supported by scarce human execution, credentialed expertise, and organizational labor. The right panel depicts the hypothesized AI-era reorganization: rents concentrate around infrastructure ownership, a narrower premium attaches to verified human presence, routine middle-tier work is compressed, and mass consumption shifts toward low-cost synthetic goods and services.}
\label{fig:economy-transitions}
\end{figure}
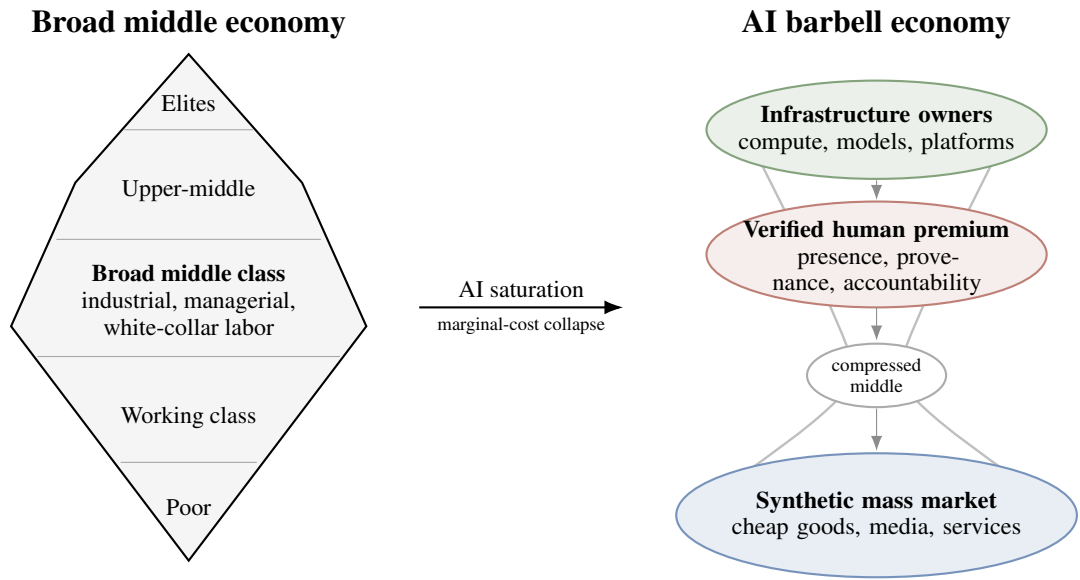

% \newpage
% \input{checklist.tex}

\end{document}